\documentclass[12pt]{article}
\usepackage{epsfig}

\begin{document}

\title{\LARGE \bf Pseudo SU(3) shell model: Normal
parity bands in odd-mass nuclei}

\author{
C. E. Vargas$^1$\thanks{Electronic address: cvargas@fis.cinvestav.mx;
fellow of CONACyT},
J. G. Hirsch$^2$\thanks{Electronic address: hirsch@nuclecu.unam.mx}, and
J. P. Draayer$^3$\thanks{Electronic address: draayer@lsu.edu}\\
{\small $^1$  Departamento de F\'{\i}sica, Centro de Investigaci\'on y de
Estudios Avanzados del IPN,}\\
{\small Apartado Postal 14-740 M\'exico 07000 DF, M\'exico}\\
{\small $^2$  Instituto de Ciencias Nucleares, Universidad Nacional
Aut\'onoma de M\'exico,}\\
{\small Apartado Postal 70-543 M\'exico 04510 DF, M\'exico }\\
{\small $^3$  Department of Physics and Astronomy, Louisiana State
University,}\\
{\small Baton Rouge, LA 70803-4001, USA }
}

\date{\today}


\maketitle

\begin{abstract}
A pseudo shell SU(3) model description of normal parity bands in $^{159}$Tb is
presented. The Hamiltonian includes  spherical Nilsson single-particle
energies,
the quadrupole-quadrupole and pairing interactions, as well as three rotor
terms.
A systematic parametrization is introduced, accompanied by a detailed
discussion
of the effect each term in the Hamiltonian has on the energy spectrum.
Yrast and
excited band wavefunctions are analyzed together with their B(E2) values.

\end{abstract}

\vskip2pc

\section{Introduction}

The shell model is a fundamental many-body approach to the study of atomic
nuclei
\cite{May49}. It explains the magic numbers as shell closures and the energy
spectra of odd-mass nuclei near closed shells as that of the odd nucleon in a
potential well defined by the closed shell nucleons. The remarkable advances in
computer power and the use of complex algorithms have allowed for systematic
studies of most $sd$- and $fp$-shell nuclei \cite{Bro88}. However, in heavy
nuclei
is not possible to solve the shell-model problem exactly. Although the
fermionic
character of the nucleons restricts their allowed degrees of freedom, the
number
of accessible states of a system still grows combinatorially with the number of
valence nucleons. For this reason truncation schemes must be introduced.

In light deformed nuclei the dominance of the quadrupole-quadrupole interaction
led to an introduction of the SU(3) shell model \cite{Ell58}. Within the SU(3)
algebraic framework, large shell-model spaces can be truncated in a very
natural
way.  While in general realistic interactions mix irreducible representations
(irreps) of SU(3), the ground state wavefunction of well-deformed light nuclei
are typically dominated by a few SU(3) irreps \cite{Aki69,Ret90,Var98}. The
strong
spin-orbit interaction renders the SU(3) truncation scheme useless in the
heavier
nuclei of the $fp$-shell, while at the same time for rare earth and actinide
species pseudo-spin emerges as a good symmetry and with it the pseudo-SU(3)
model
\cite{Hech69,Ari69,Rat73}.

Pseudo-spin can be recognized from the experimental fact that single-particle
orbitals with $j = l - 1/2$ and $j = (l-2) + 1/2$ in the shell $\eta$ lie very
close in energy and can therefore be labeled as pseudo spin doublets with
quantum
numbers $\tilde{\j} = j$, $\tilde\eta = \eta -1$, and $\tilde l = l - 1$. The
origin
of this symmetry has been traced back to the relativistic mean field equations
\cite{Blo95,Gin97,Men98}.

In this work the energy spectra and B(E2) transition strengths of
$^{159}{Tb}$ are
calculated from a microscopic perspective using the pseudo SU(3) shell model.
Normal parity bands are described in a many-body basis built with active
nucleons
occupying normal parity levels. Polarization effects due to valence nucleons in
intruder orbits are taken into account through the use of effective charges.

Basis states are built from SU(3) irreps obtained by taking the direct
product of
proton and neutron representations. In \cite{Zuk95,Var98} it is shown that
for a
description of the low-energy spectrum of deformed nuclei, the Hilbert space
should be truncated according to contributions of the  quadrupole-quadrupole
interaction and the single-particle one-body Hamiltonian. While pairing plays a
very important role in determining the energy spectrum, but does not strongly
modify the wave functions of deformed nuclei.

As can be seen in \cite{Beu98,Var99}, by using an schematic Hamiltonian
parametrized according to systematics \cite{Rin79,Duf96}, it is possible to
describe the low-lying energy spectrum of even- and odd-mass  heavy deformed
nuclei. While the application of the model to other deformed rare-earth and
actinide nuclei is in order, in the present contribution we will address some
specific questions about the pseudo SU(3) model. In particular, we will discuss
the importance of various terms in the Hamiltonian, ways in which their
strengths
can be deduced from systematic, and the effect each term has on the energy
spectra.
Taking $^{159}{Tb}$ as an example, we will examine the wave functions of each
rotational band, paying particular attention to the $B \{ E2; J \rightarrow
(J-1)
\}$ and $B \{ E2; J \rightarrow (J-2) \}$ transition strengths between
states in
the same bands and between states belonging to different bands.

In section 2 the pseudo SU(3) classification scheme is presented. The
pseudo SU(3)
Hamiltonian and its parametrization is discussed in Section 3. The effect which
each term in the Hamiltonian has on the energy spectra is analyzed in
Section 4.
Section 5 contains the analysis of the wave functions associated with different
rotational bands and the associated B(E2) values. Conclusions are discussed in
Section 6.

\section{The pseudo SU(3) basis}

The first step in any application of the pseudo SU(3) model is to build the
many-body basis. To do this it is necessary to know how many valence nucleons
occupy the normal parity orbitals. We will show how this is done using
$^{159}$Tb
as an example. It has 65 protons and 94 neutrons, and of these, 15 protons
and 12
neutrons are in the last unfilled (open) shells.  Assuming a deformation $\beta
\sim 0.25 $, the deformed Nilsson single-particle levels of the active
shells are
filled from below \cite{Dra84,Cas87}. Nine protons are distributed in the
$1g_{7/2}$ and $2d_{5/2}$ orbitals of the $\eta = 4$ shell, and the
remaining six
occupy the $1h_{11/2}$ intruder orbital. Eight neutrons occupy the
$2f_{7/2}$ and
$1h_{9/2}$ orbitals of the $\eta = 5$ shell and four are in $1i_{13/2}$
orbital.
The relevant occupation numbers $n_\pi, n_\nu$ can be summarized as

\begin{eqnarray}
n^N_\pi = 9, ~~~ n^A_\pi = 6, ~~~ n^N_\nu = 8, ~~~ n^A_\nu = 4
\label{eq:ocup} \end{eqnarray}
where $N$ refers to normal parity states and $A$ to the abnormal parity
(also called unique or intruder) states. The deformed Nilsson mean field is
only employed to define the number of nucleons in normal and unique parity
orbitals. The use of the pseudo SU(3) basis to describe the normal parity
sector implies that these nucleons can occupy all normal parity orbitals,
not only those with the lower single particle energies.

As it has been the case for all pseudo SU(3) studies up to now, we will
freeze the
nucleons in abnormal parity orbital and describe the dynamics using only
nucleons
in normal parity states. While it has been shown that this is a reasonable,
it is
nonetheless a strong assumption. This choice is further reflected through
the use
of effective charges to describe quadrupole electromagnetic transitions
which are
larger than those usually employed in typical shell-model calculations for
light
nuclei. A more sophisticated treatment of the problem, with nucleons in
intruder
orbitals described in the same footing using SU(3) irreps is under development
\cite{Var98}.

The many-particle states of $n_\alpha$ active nucleons in a given normal parity
shell $\eta_\alpha$, $\alpha = \nu$ or $\pi$ can be classified by the following
chains of groups:

\begin{eqnarray}
~\{ 1^{n^{N}_\alpha} \} ~~~~~~~ \{ \tilde{f}_\alpha \} ~~~~~~~~\{ f_\alpha
\} ~\gamma_\alpha ~ (\lambda_\alpha , \mu_\alpha ) ~~~ \tilde{S}_\alpha
~~ \kappa_\alpha  \nonumber \\
~U(\Omega^N_\alpha ) \supset U(\Omega^N_\alpha / 2 ) \times U(2) \supset
SU(3) \times SU(2) \supset \nonumber \\
\tilde{L}_\alpha  ~~~~~~~~~~~~~~~~~~~~~ J^N_\alpha ~~~~ \nonumber \\
SO(3) \times SU(2) \supset SU_J(2),
\label{eq:chains}
\end{eqnarray}

\noindent where above each group the quantum numbers that characterize its
irreps
are given and $\gamma_\alpha$ and $\kappa_\alpha$ are multiplicity labels
of the
indicated reductions.

Any state $| J_i M \rangle$, where $J$ is the total angular momentum, $M$ its
projection and $i$ an integer index which enumerates the states with the same
$J, M$ starting from the one with the lowest energy, is built as a linear
combination
\begin{equation}
| J_i M \rangle = \sum_\beta C^{JMi}_\beta |\beta JM \rangle \label{wf}
\end{equation}
of the strong coupled proton-neutron states
\begin{eqnarray}
|\beta JM \rangle \equiv
| \{ \tilde{f}_\pi \} (\lambda_\pi \mu_\pi) S_\pi, \{ \tilde{f}_\nu \}
(\lambda_\nu \mu_\nu) S_\nu ; \rho (\lambda \mu ) \kappa L,S~ JM \rangle
~~~~~~~~~~~~~~~~~~~~~~~~~~~~
\nonumber \\
 =   \sum_{M_L M_S} (L M_L, S M_S | J M )
\sum_{M_{S \pi} M_{S \nu}}
(S_\pi M_{S \pi}, S_\nu M_{S \nu} | S M_S)  \nonumber \\
\sum_{k_\pi \kappa_\nu L_\pi L_\nu M_\pi M_\nu }
{ \langle (\lambda_\pi \mu_\pi) \kappa_\pi L_\pi M_\pi ;
(\lambda_\nu \mu_\nu) \kappa_\nu L_\nu M_\nu |
(\lambda \mu ) \kappa L M \rangle}_\rho \label{basis} \\
| \{ \tilde{f}_\pi \} (\lambda_\pi \mu_\pi) \kappa_\pi L_\pi M_\pi, S_\pi M_{S
\pi}  \rangle
| \{ \tilde{f}_\nu \} (\lambda_\nu \mu_\nu) \kappa_\nu L_\nu M_\nu, S_\nu M_{S
\nu} \rangle  .\nonumber
\end{eqnarray}

In the above expression $\langle-;-|- \rangle$ and $(-,-|-)$ are the SU(3) and
SU(2) Clebsch Gordan coefficients, respectively.  We are considering only
configurations with the highest spatial symmetry \cite{Var98,Dra82}. For
$^{159}$Tb the active shells in the pseudo SU(3) space are $\tilde\eta_\pi
=3$ and
$\tilde\eta_\nu = 4$ with degeneracies $\Omega_\pi = 20$ and $\Omega_\nu = 30$,
respectively. In the {\em large} groups $U(10)$ and $U(15)$, the spatially most
symmetric irreps for 9 protons and 8 neutrons are, respectively, $ \{
\tilde{f}_\pi \} = \{ 2^4 1 \}$ and $\{ \tilde{f}_\nu \} = \{ 2^4 \} $. It
implies
that $\tilde{S}_{\pi} = 1/2$ and $\tilde{S}_{\nu} = 0$. In other words, we
are only
taking into account configurations with pseudo spin zero for an even number of
nucleons and $1/2$ for an odd number of nucleons.

The above considerations rely strongly on the goodness of pseudo-spin symmetry,
which manifests itself in the near degeneracy of the pseudo spin-orbit
partners.
When the {\em real} SU(3) model is used to describe deformed nuclei in the {\em
pf}-shell, the spin mixing is very important and Eq.  (\ref{basis}) must be
modified accordingly \cite{Var98,Hir99}.

What makes the pseudo SU(3) model a powerful theory is that it allows one
to envoke
a relatively simple and physically motivated basis truncation scheme. Extended
shell-model calculations in the {\em pf}- and {\em sdg}-shell have shown
that in
the description of deformed nuclei the Hilbert space can be truncated to only
those states that are relevant when both the quadrupole-quadrupole force
and the
single-particle Hamiltonian are taken into account \cite{Zuk95}. While
pairing is
fundamental to obtaining the correct moment of inertia of the rotational
bands, it
has a relatively small effect on the overall wave functions \cite{Zuk95}. An
analysis of the SU(3) content of wave functions obtained in large shell-model
diagonalizations \cite{Hir99}, as well as the excellent description of
ground and
excited bands in heavy deformed even- \cite{Beu98, Beu99} and odd-mass
\cite{Var99} nuclei strongly support this statement.

The quadrupole-quadrupole interaction can be expressed in terms of the
second order SU(3) Casimir operator $C_2$,

\begin{equation}
\hat{Q} \cdot \hat{Q} = 4 C_2 - 3 \hat{L}(\hat{L}+1).
\label{eq:qq} \end{equation}
The  eigenvalue of $C_2$ for a given of SU(3) irrep
$(\lambda,\mu)$ is given by
\begin{equation}
\langle C_2 \rangle = ({\lambda}^2 + {\mu}^2 + \lambda \mu + 3 \lambda +
3 \mu ) .
\end{equation}
The larger the expectation value of $C_2$, the greater the binding of that
SU(3)
irrep by a pure $Q\cdot Q$ interaction. We build the basis selecting the proton
and neutron irreps with the largest $ \langle C_2 \rangle$.

For $^{159}{Tb}$, the SU(3) representations were selected as follows.
Including all
the possible spins there are 97 irreps in the proton space and 285 irreps
in the
neutron space. Proton and neutron irreps which belongs to the irreps $\{
\hat{f}_\pi \} = \{ 2^4 1 \}$  and $\{\hat{f}_\nu \} = \{ 2^4  \}$ of the large
groups $U(10)$ and $U(15)$, respectively, are found and ordered according
to their
$C_2$ value. In Table 1 the seven representations with the largest $C_2$ in
$^{159}{Tb}$ are shown.

Table 1

Taking the direct product of these protons and neutrons irreps results in many
strong coupled SU(3) irreps. From these, the 15 with the largest $C_2$ were
chosen. They are shown in  Table 2. These 15 proton-neutron irreps define the
Hilbert space of the model. They are a small subset of all the possible
irreps, and
involve only 4 proton and 3 neutron irreps. In this strongly truncated
space it is
possible to describe the low energy spectra of even-even nuclei
\cite{Beu98,Beu99}
and normal parity bands in odd-mass nuclei \cite{Var99}. The validity of
the pseudo
SU(3) symmetry is the rationale behind this successful truncation scheme.

Table 2.

\section{The pseudo SU(3) Hamiltonian}

The Hamiltonian contains spherical Nilsson single-particle terms for protons
($H_{sp,\pi}$) and neutrons ($H_{sp,\nu}$), the quadrupole-quadrupole
($\tilde Q \cdot \tilde Q$) and pairing interactions пн($H_{pair,\pi}$ and
$H_{pair,\nu}$), as well as three `rotor-like' terms which are diagonal in the
SU(3) basis.

\begin{eqnarray}
 H & = & \sum_{\alpha =\pi, \nu} \{ H_{sp,\alpha} - ~ G_\alpha
~H_{pair,\alpha} \}
         - \frac{1}{2}~ \chi~ \tilde Q \cdot \tilde Q \label{ham} \\
   &   & + ~a~ K_J^2~ +~ b~ J^2~ +~ A_{asym}~
         \hat C_2 .\nonumber
\end{eqnarray}

This Hamiltonian can be separated into two parts: the first row includes
Nilsson
single-particle energies and the pairing and quadrupole-quadrupole
interactions ($\tilde{Q}$ is the quadrupole operator in the pseudo SU(3)
space, see below).
They are the basic components of any realistic Hamiltonian \cite{Rin79,Duf96}
and have been widely studied in the nuclear physics literature, allowing
their respective strengths to be fixed by systematics \cite{Rin79,Duf96}.
In the second row there are three rotor terms used to fine tune the moment of
inertia and the position of the different $K$ bands. The SU(3) mixing is due to
the single-particle and pairing terms.

The three `rotor-like' terms have been studied in detail in previous papers
where
the pseudo SU(3) symmetry was used as a dynamical symmetry
\cite{Dra84,Cas87}. In
the present work, $a$ and $b$ are the only two parameters used to fit the
spectra.

The term proportional to $K_J^2$ breaks the SU(3) degeneracy of the different K
bands \cite{Naq90}. It has the form

\begin{equation}
K^2_J = \frac{\lambda_1 \lambda_2 J^2 + \lambda_3 X^c_3 + X^c_4}
{2 \lambda^c_3 + \lambda_1 \lambda_2 }
\end{equation}

\noindent where $J^2$, $X^c_3$ and $X^c_4$ are the three rotational scalars
formed with products of $J$ and $\tilde Q$ and the $\lambda_i$
coefficients are functions of $\lambda$ and $\mu$
\cite{Dra82,Dra84,Cas87}, as shown in Appendix A.
The reduced matrix elements for $X^c_3$ and $X^c_4$
are evaluated using Racah and SU(3) coupling coefficients \cite{Naq90,Tho98}.
For $^{159}$Tb $a = 0.0198$ was found to provide the best fit.

The term proportional to $J^2$ is used to fine tune the moment of inertia.
It represents a small correction to the  quadrupole-quadrupole term,
which contributes to the rotor spectra with strength $3/2 \chi$ (see Eq.
(\ref{eq:qq})). For $^{159}$Tb we used $b = -0.0031$, which introduce a
change of about 15\% in the rotational spectra.

The asymmetry term distinguishes SU(3) irreps with both $\lambda$ and $\mu$ even
from the others \cite{Les87}, having no interaction strength in the first case and
a positive one in the the second. In this way the contribution of irreps
with both
$\lambda$ and $\mu$ even is slightly enhanced because they belong to different
symmetry types of the intrinsic Vierergruppe $D_2$ \cite{Les87}. The asymmetry
coefficient has a value $A_{asym} = 0.0008$, fixed according with \cite{Tho98}.
The same value was employed for the three A=159  nuclei studied in
\cite{Var99}.

The single-particle Nilsson Hamiltonian is

\begin{equation}
H_{sp} = \hbar \omega_0 (\eta + \frac{3}{2}) - \kappa \hbar {\omega}_0 \{ 2
\vec{l} \cdot \vec{s} + \mu \vec{l}^2 \},
\label{Nilssonh} \end{equation}

\noindent with parameters \cite{Rin79}

\begin{eqnarray}
\hbar {\omega}_0 = 41 A^{-1/3} [MeV],
&\kappa_\pi =  0.0637, & \kappa_\nu = 0.0637, \\
&\mu_\pi = 0.60, &\mu_\nu = 0.42, \nonumber
\end{eqnarray}

The pairing interaction is

\begin{equation}
V_p = -\frac{1}{4} G \sum_{j,j'} a^\dagger_j
a^\dagger_{\bar{j}} a_{j'} a_{\bar{j'}} \label{pair}
\end{equation}

\noindent where $\bar{\j}$ denotes the time reversed
partner of the single-particle state $j$ and
$G$ is the strength of the pairing force. Its second quantized expression
in term of SU(3) tensors is reviewed in Appendix B. For the pairing
coefficients
$G_{\pi,\nu}$, we used \cite{Rin79,Duf96}

\begin{equation}
G_\pi = \frac{21}{A} = 0.132, ~~~~~~~~~
G_\nu = \frac{17}{A} = 0.106  .
\end{equation}

In the SU(3) model the collective quadrupole operator, defined by
$Q^c_\mu = \sqrt{ 16 \pi/5} \sum_i r_i^2 Y_{2\mu}(\hat r_i)/b^2$, is
symmetrized in order to correspond to one of the SU(3) generators. It is
called the `algebraic' quadrupole operator $Q^a_\mu = \sqrt{ 4 \pi/5}
\sum_i [r_i^2 Y_{2\mu}(\hat r_i)/b^2 + b^2 p_i^2 Y_{2\mu}(\hat p_i)]$
\cite{Dra84,Cas87}.  Within a major oscillator shell, the matrix elements
of $Q^c$
and $Q^a$ are identical. When transformed to the pseudo SU(3) basis, it
maps to a
linear combination of SU(3) tensors which is dominated by the quadrupole
operator
$\tilde Q$, which is the generator of the pseudo SU(3) algebra
\cite{Cas87}. It is
this quadrupole operator in the pseudo SU(3) space the one included in the
quadrupole-quadrupole interaction in Hamiltonian (\ref{ham}). In first
order, the
relationship between both quadrupole operators can be written as

\begin{equation}
 Q^a_\mu =  {\frac {\tilde\eta +1} {\tilde\eta}} \tilde Q_\mu
\end{equation}
which holds for protons and neutrons separately.

The coefficient $\chi$ of the operator $\tilde Q \cdot \tilde Q$ is
\begin{equation}
\chi = {\frac {35} {A^{5/3}} } = 0.00753 .
\end{equation}
It is consistent with the parametrization discussed in \cite{Duf96},
provided one keep in mind that in this reference the quadrupole
operator is just $r^2 Y_{2\mu}$, and that, as mentioned above, when
operating in the pseudo SU(3) space the interaction must be, in first
order, a factor $\left( {\frac {\tilde\eta +1} {\tilde\eta}}\right)^2$ stronger
than
the similar one in the normal space. This implies that the constant $\chi$
in \cite{Duf96} is a factor $ {\frac {16 \pi /5 }
{[(\tilde\eta+1)/\tilde\eta ]^2}} \approx 6.5$ larger than our $\chi$.

The electric quadrupole operator is expressed as\cite{Dra82}

\begin{equation}
Q_\mu = e_\pi Q_\pi + e_\nu Q_\nu \approx
e_\pi {\frac {\eta_\pi +1} {\eta_\pi}} \tilde Q_\pi +
e_\nu {\frac {\eta_\nu +1} {\eta_\nu}} \tilde Q_\nu , \label{q}
\end{equation}

\noindent
with effective charges $e_\pi = 2.3, ~e_\nu = 1.3$. These values
are very similar to those used in the pseudo SU(3) description  of
even-even nuclei \cite{Dra82,Beu98}. They are larger than those used in
standard calculations of B(E2) strengths \cite{Rin79} due to the passive
role assigned to the nucleons in unique parity orbitals, whose
contribution to the quadrupole moments is parametrized in this way.

\section{The energy spectra}

Fig. 1 shows the yrast and excited bands in $^{159}$Tb. Experimental data
\cite{nndc} are plotted on the left hand side, while those obtained using
the Hilbert space and the Hamiltonian parameters discussed in the
previous sections are shown in the right hand side. The agreement between
both is excellent, as is the case for the nuclei $^{159}$Eu and $^{159}$Dy,
whose energy spectra was studied in \cite{Var99}. From the four bands
reported in
the literature in $^{159}$Tb, three of them (the yrast, ${5/2}^+_1$, and
${3/2}^+_2$ bands) have a difference between the experimental and
predicted
levels of less than 50 KeV. The ${1/2}^+_1$ is slightly high in
energy, and
the model predicts an exaggerated staggering, with origin discussed below.

Figure 1

The whole energy spectra is built up by the interplay between the
single-particle
and quadrupole-quadrupole terms in the Hamiltonian \cite{Zuk95,Var98}.
These two
terms define the relative ordering between the different bands, as well as
the main
components of the wavefunction. As expected, the use of realistic
single-particle
energies plays a key role in the appropriate description of odd-mass nuclei.

To make this point clear, in Fig. 2 the theoretical energy spectra
calculated with
Hamiltonian (\ref{ham}) {\em without single-particle energies} are
presented on the
right of each column and compare it with the corresponding experimental
energies on
the left of each column \cite{nndc}. It is clear that the ordering is
shifted. The
ground state is now predicted to have $J = {\frac 1 2}$, the first excited band
starts with $J = {\frac 5 2}$. Only the third band with $J = {\frac 3 2}$
reamains
ordered correctly. The wave functions of all the states in all these three
bands are
dominated (more than 95\%) by the leading SU(3) irrep $(28,8), (10,4)_\pi
(18,4)_\nu$. We will return to this point when discussing the wave functions in
Section 5.

Figure 2

From Fig. 2 it is also clear that, in absence of single-particle energies, the
quadrupole-quadrupole interaction rules over all the others, even in the
presence
of realistic pairing strengths. It determines the single SU(3) irrep that
dominates
the low energy spectra and the rotor pattern each band exhibits.

In order to discuss the effect each of the remaining term in Hamiltonian
(\ref{ham}) has on the energy spectra, three rotational bands are plotted in
Figure 3: the yrast band in insert (a), the ${5/2}^+_1$ band in
insert (b)
and the ${1/2}^+_1$ band in insert (c). In the first column from the
left the
experimental energies are plotted for each band, with the angular momentum and
parity of each state written on the left. The next column presents the
theoretical
results, the same ones shown in Fig. 1. The next three columns show the
theoretical
spectra obtained {\em without} the pairing interaction (third column, $G_\pi =
G_\nu = 0$), without pairing between protons (fourth column, $G_\pi = 0$) and
without pairing between neutrons (fifth column, $G_\nu = 0$). The remaining
four
columns depict the behavior of the spectra when the rotor terms are turned off.
The sixth column, labeled $K^2$, shows the spectra without this term
($a=0$), the
seventh column without the $J^2$ term ($b =0$), the eighth without the
asymmetry
term ($A_{asym} = 0$) and the ninth presents the spectra with all the three
rotor
terms turned off ($a = b = A_{asym} = 0$). The last column on the right
presents
the experimental energies again, to help with the comparison.

Figure 3

The theoretical description of the yrast band is good. The effect of the
pairing
interaction is clearly seen in the third column; namely, it expands the energy
spectra. This is a rather remarkable result that seems to comes about
because of
the highly truncated nature of the basis. The pairing interaction strongly
mixes
the SU(3) irreps \cite{Tro95,Bah95,Esch98,Bah98} and plays an important role in
determining the moment of inertia of this deformed nuclei. At the same time its
effects on the wavefunction are minor, as has been discovered in indenpendent
shell-model calculations \cite{Zuk95}. Comparing the third, fourth and fifth
columns, it is clear it is the pairing interaction between neutrons that is
most
important. The last uncoupled proton in $^{159}$Tb seems to nearly ignores the
effect of pairing. The $K^2$ term has negligible effect on the yrast band. The
$J^2$ term helps to fine tuning the moment of inertia to the experimental
number.
The asymmetry term pushes the $J= {\frac {17} 2}$ state down in energy and
closer
to the experimental value. Although a small contribution it is clearly
important
and provides justification for including this term included in the
Hamiltonian.

The ${5/2}_1$ band shown in insert (b) shows a similar behavior. The first
three
states reported are in close agreement with the experimental data, and theory
predicts more states that belongs to this band. In the absence of pairing, the
energy spectra is compressed which is due mainly to the protons. The $K^2$
term is
clearly relevant for this band: when it is missing the band head moves down
about
100 KeV in energy. The $J^2$ term modifies the moment of inertia slightly.

The ${1/2}_1$ band is shown in insert (c). Its band head is predicted at an
energy
150 KeV larger than the measured one. The theory predicts three nearly
degenerated
pairs of levels: $({\frac 3 2}, {\frac 5 2}), ({\frac 7 2}, {\frac 9 2})$ and
$({\frac {11} 2},{\frac {13} 2})$ while their experimental counterparts are
less
closely packed. As can be seen in the eighth column, when $A_{asym} = 0$
the level
spacing is closer to the reported one, but the band-head energy it too
high. It is
noticeable than turning off the $K^2$ term the energies move upwards, while
in the
${5/2}_1$ band they move downward.

The contribution of each term in the Hamiltonian can be summarized as follows:

\begin{itemize}
\item{\bf Quadrupole-quadrupole}: This interaction is tied to the quadrupole
deformation, and since the pioneering work of Elliott \cite{Ell58} has been
known
to play a crucial role in the dynamics of deformed nuclei. In the pseudo
SU(3) model
its dominance can also be used to determine the truncation of the Hilbert
space.

\item{\bf Spherical single-particle energies}: These form the basis of the
shell
model, and serve define the low-energy spectra of odd-mass nuclei. And this is
precisely the role they play in the pseudo SU(3) model: they determine to a
large
extent relative ordering of the different bands. Together with the
quadrupole-quadrupole interaction, the single-particle energies define the
gross
features of the energy spectra and the mixing of different SU(3) irreps in
the wave
functions.

\item{\bf Pairing}: This interaction expands the whole energy spectra,
almost as
though it is a multiplicative constant in front of Hamiltonian (\ref{ham}).
It also
shifts the energy of each band head and alters the moment of inertia of
each band.
The success of previous investigations that used the pseudo SU(3) symmetry as a
dynamical symmetry \cite{Dra84,Cas87} without the mixing of SU(3)
representations,
is partially a result of this effect: a larger quadrupole-quadrupole and rotor
strength mimics the effect of pairing on the spectra. The pairing interaction
acting in the subspace with an odd number of nucleons has a nearly negligible
effect on the low-lying bands. The subspace with even number of nucleons is
responsible for most of the pairing effects.

\item{$\mathbf K^2$}: Since it has been shown that this particular
combination of
2-, 3- and 4-body terms corresponds to the square of the third component of the
angular momentum in the intrinsic frame \cite{Les87}, it can be used to adjust
band-head energies. For the ${5/2}_1$ band, when $a=0$ the band-head energy
is 277 keV, while in the presence of this term it moves up to 356 keV,
which is
close to the experimental value of 348 keV \cite{nndc}.

\item{$\mathbf J^2$}: This diagonal term provides small corrections to the
moment
of inertia. The negative $b$ value used in the present study makes serves to
increase the moment of inertia, compressing the corresponding spectra.

\item{\bf Asymmetry}: This term enhances the contribution of the SU(3) irreps
with both $\lambda$ and $\mu$ even. It has an important effect on states with
large angular momentum. For example, in the absence of this term the state
${\frac
{17} 2}^+$ of the yrast band is displaced to higher energies due to mixing with
other SU(3) irreps with $\lambda$ or $\mu$ odd.

\item{\bf Rotor terms}: The effect of suppressing simultaneously the three
rotor
terms, {\em i.e.} taking $a = b = A_{asym} = 0$, is shown in the ninth
column of
Fig. 3. The inclusion of Nilsson single-particle energies and the pairing and
quadrupole-quadrupole interactions suffices to provide a very reasonable
energy
spectra, with all known bands in their correct order and the overall rotor
features
reproduced. The rotor terms provide the fine tuning, with energies being
adjusted
by no more than 15 \%. The predictive power of the pseudo SU(3) model strongly
relies in this fact.

\end{itemize}

\section{B(E2) transition strengths and wave functions}

Up to this point we have centered the discussion on the energetics. But the
pseudo
SU(3) model is far more powerful than this, it can also successfully
describe the
electromagnetic transitions. Most of the B(E2) transition strengths between
states
in the first four bands in $^{159}$Tb are presented in Fig. 4 for the
$J\rightarrow
J-1$ transitions, and in Fig. 5 for the $J \rightarrow J-2$ transitions.

Figure 4

Figure 5

Calculated B(E2) values are given in units of $e^2 b^2 \times 10^{-2}$.
They are written close to the arrow which graphically show the transition they
describe. The effective charge employed was discussed in relation with Eq.
(\ref{q}). This is the only extra parameter introduced, and has a value
very close
to those used in previous studies \cite{Cas87}.  To assess the quality of the
results, the experimental B(E2) values \cite{nndc} between yrast band
states are
reported in Table 3. Most of the transition strengths are reproduced within the
experimental error bars. One exception is the transition
${17/2}_1 \rightarrow {13/2}_1$ which is underestimated.
This could be related with the change in the wavefunction of the first ${\frac {17}
2}$, whose mixing with the second ${\frac {17} 2}$ state seems to be exaggerated in
the model, as discussed below.

Table 3

Notice that while the intraband B(E2) transition strengths are on the order of
hundreds $e^2 b^2 \times 10^{-2}$, the interband transitions are much less,
typically on the order of $e^2 b^2 \times 10^{-2}$ or fractions thereof. This fact
supports the
identification of states belonging to bands, and is consistent with the
wavefunction analysis. The only measured interband transition is the
${1/2}_1 \rightarrow {5/2}_1$ and it too is well reproduced by the model,
wich predicts a value of 2.96 $e^2 b^2 \times 10^{-2}$.

It is very instructive to analyze the wave functions of the states
belonging to the
lowest lying energy bands obtained with the Hamiltonian (\ref{ham}). The
nine most
important SU(3) irreps which are relevant to the description of these bands are
listed in Table 4.

Table 4

In Figure 6 the percentage each irrep contributes to each state is plotted as a
function of the angular momentum of the state for the different bands.
These were
calculated from the wavefunctions, Eq. (\ref{wf}), as simply $100 \times
|C^{JMi}_\beta|^2$. The symbols listed in the first column of Table 4
identifies
the various components. All contributions larger than 2 \% are plotted, and
in all
cases the states shown add up to at least 95\% of the total wavefunction.

Figure 6

Each insert, (a) to (d) in Figure 6, gives the main components of one of the
bands plotted in Figs. 4 and 5, from the band head and up to the state with $J=
{\frac {15} 2}$ or ${\frac {17} 2}$. As shown, all have a very regular
structure
as one moves up the bands. While in all the cases there is strong mixing of
SU(3)
irreps, the mixing remains nearly the same for the states with different
angular
moments belonging to the same band. In this sense the mixing is $adiabatic$
within
each band. This coherence explains the large B(E2) values for intra-band
transitions.

In insert (a) the components of each state belonging to the ${3/2}_1$ band are
shown. As shown, the states from $J^\pi$~=~${3/2}^+$ to ${17/2}^+$ are about
30 \%  {$(30,4) {[(10,4)}_\pi \times {(20,0)}_\nu$]}, 29 \% $(28,8)
{[(10,4)}_\pi
\times {(18,4)}_\nu$], 14 \% $(30,4) [{(11,2)}_\pi \times {(18,4)}_\nu$], 12 \%
$(31,2) {[(11,2)}_\pi \times {(20,0)}_\nu$] with smaller contributions from the
$(31,2)$ and $(29,6)$ representations that are almost constant for all the
states
in the band.

States in the ${5/2}_1$ band, insert (b), have about 36 \% $(28,8)
{[(10,4)}_\pi \times {(18,4)}_\nu$], 22 \% $(30,4) {[(11,2)}_\pi \times
{(18,4)}_\nu$], and 20 \% $(30,4) {[(10,4)}_\pi \times {(20,0)}_\nu]$ which
decreases to less then 15\% for the $J= {\frac {15} 2}$ state. Other irreps
contribute less than 10\% each. The ${1/2}_1$ band, insert (c), is the
purest of
the four considered.  It has around 60 \% $(28,8) {[(10,4)}_\pi \times
{(18,4)}_\nu$], 20\%  $(30,4) {[(11,2)}_\pi \times {(18,4)}_\nu$], and 10 =\%
$(30,4) {[(10,4)}_\pi \times {(20,0)}_\nu$]. It is interesting to note that
this
band becomes the ground state band when the single-particle energies are not
present (see the discussion  below Figure 2), and in that case all the
states in
the four bands are built primarily out of the $(28,8) {[(10,4)}_\pi \times
{(18,4)}_\nu$] irrep.

The ${3/2}_2$ band, insert (d), is dominated by the irrep $(30,4) {[(10,4)}_\pi
\times {(18,4)}_\nu$] which get strongly mixed with the $(28,8) {[(10,4)}_\pi
\times {(18,4)}_\nu$] for $J = {\frac {11} 2}$ and  ${\frac {13} 2}$.

As was mentioned above, the interplay between the single-particle and the
quadrupole-quadrupole terms in the Hamiltonian defines the SU(3) mixing in
the wave
functions. The four bands discussed here have strong mixing, and the ground
state
band is {\em not} dominated by the leading irrep, the one which would
constitute
the ground band in the pure SU(3) symmetry limit. The delicate balance between
these two interactions, whose strengths are taken from known systematics
and not
used as fitting parameters, defines the gross features of the calculated energy
spectra which are found to be in good agreement with the available experimental
information.

Although the basis is strongly truncated, being built from the 15 SU(3)
irreps listed in Table 2, not even all of these play an important role. The
low-lying energy bands discussed above are dominated by the irreps coming
from the first four rows in Table 2, which are combinations of the two proton
SU(3) irreps ${(10,4)}_\pi$ and ${(11,2)}_\pi$ and the two neutron SU(3) irreps
${(18,4)}_\nu$ and ${(20,0)}_\nu$.

\section{Conclusions}

We have shown that the normal parity bands in the odd-mass heavy deformed
nuclei with A = 159 can be described quantitatively using the pseudo SU(3) model
\cite{Var99}. A careful examination of the wave functions and B(E2) transition 
strengths for the four low-lying energy bands in $^{159}{Tb}$ was made, analyzing 
their structure in terms of their SU(3) components, and relating them with their 
intra- and inter-band transitions,
for which the calculated values agree closely with the known experimental
numbers.

The most relevant feature of the present application of the pseudo SU(3)
model is
a determination of the primary features of the energy spectra of the normal
parity
rotational bands in $^{159}$Tb using a Hamiltonian with Nilsson
single-particle energies and quadrupole-quadrupole and pairing interactions
with
strengths fixed by systematics -- strengths of the primary interactions
were $not$
varied to obtain a ``best fit'' to the date. A few extra rotor-like terms
were used
to obtain a more precise description of the energies and B(E2) values, but this
``fine tuning'' not affect the spectra in a major way and had little
influence on
the structure of the calculated wavefunctions.

This work shows the pseudo SU(3) model to be a powerful shell-model scheme, one
that can be used to describe normal parity bands in deformed rare-earth and
actinide
isotopes by performing a symmetry dictated truncation of the Hilbert space and
using a systematic parametrization of the dominant terms in the Hamiltonian. It
opens up the possibility of a detailed microscopic analysis of other nuclear
properties of heavy deformed nuclei, both with even and odd numbers of
protons and
neutrons, like g-factors, M1 transitions and beta decays.

\section{Acknowledgments}
This work was supported in part by Conacyt (M\'exico) and the U.S. National
Science
Foundation, the latter being through Grants 9500474 and 9603006, as well as
an NSF
Cooperative Agreement, 9720652, that includes matching from the Louisiana Board
of Regents Support Fund.

\bigskip

\appendix{\bf Appendix A: The $K_J^2$ term}

\bigskip

The residual interaction, $K^2$, is a linear combination of  ${J}^{2}$,
${X}_{3}$ and ${X}_{4}$, which are rotational scalar operators built from
generators of the $SU(3)$ algebra, that is \cite{Dra84,Cas87},
\begin{eqnarray}
{J}^{2} = & \sum_{i}^{3}{{J}_{i}^2}\nonumber \\
                 {X}_{3} = & \sum_{i,j}^{3}{{J}_{i}
                {Q}_{ij}^a {J}_{j}} \\
                {X}_{4} = & \sum_{i,j,k}^{3}{{J}_{i}
                {Q}_{ij}^a {Q}_{jk}^a {J}_{k}} \nonumber \\
\end{eqnarray}

\noindent
where ${J}_{i}$ and ${Q}_{ij}^a$ are cartesian forms of the total angular
momentum and the quadrupole operators, respectively. The $K$ is interpreted
to be
the third component of the total angular momentum along the intrinsic
body-fixed
symmetry axis of the system, which is given by \cite{Dra84,Cas87}

\begin{equation}
                {K}^{2} = ({ \lambda}_{1} { \lambda}_{2} {J}^{2} +
                 { \lambda}_{3} {X}_{3} + {X}_{4}) / ( 2
                 { \lambda}_{3}^2 + { \lambda}_{1}
                 { \lambda}_{2}) \ , \label{K2}
\end{equation}
with the parameters $\lambda_i$ denoting the eigenvalues of the mass
quadrupole operator, which are related to the $SU(3)$ labels
$(\lambda,\mu)$ through the expressions \cite{Cas88,Naq90}.

\begin{equation}
                \lambda_1 = {1 \over 3} (\mu-\lambda) , \ \
                \lambda_2 = -{1 \over 3} (\lambda + 2 \mu + 3), \ \
                \lambda_3 = {1 \over 3} (2 \lambda +  \mu + 3) .
\end{equation}

The last expressions can be obtained by requiring a linear correspondence
between
the invariants of the $SU(3)$ and the semidirect product $T_5 \wedge SO(3)$
groups
\cite{Cas88,Naq90}. Indeed, the expectation value of the operator
(\ref{K2}), with
respect to an orthonormalized basis associated to the chain of groups
$SU(3) \rightarrow SO(3)$, corresponds to the eigenvalues of the third
component of an intrinsic angular momentum when $L \ll min (\lambda,\mu)$.

\bigskip

\appendix{\bf Appendix B: The pairing interaction}

\bigskip

The pairing term (\ref{pair}) in the Hamiltonian can be expressed in
second quantization as \cite{Tro95,Bah95}

\begin{eqnarray}
V_p = \frac{G}{2} \sum_{(\lambda_1,\mu_1)(\lambda_2,\mu_2)} \sum_{\eta
\eta '}
P_{\eta \eta '} \{ (\lambda_1,\mu_1)(\lambda_2,\mu_2) \rho_0
(\lambda_0,\mu_0) \} \nonumber \\
{[ {[ a^\dagger_\eta \otimes a^\dagger_\eta ]}^{\lambda_1 \mu_1} \otimes
{[
\tilde{a}_{\eta '} \otimes \tilde{a}_{\eta '} ]}^{\lambda_2 \mu_2}
]}^{\rho_0
\lambda_0 \mu_0 k_0 = 1 l_0 = s_0 = 0}
\end{eqnarray}

\noindent where

\begin{eqnarray}
P_{\eta \eta '} \{ (\lambda_1,\mu_1)(\lambda_2,\mu_2) \rho_0
(\lambda_0,\mu_0) = \nonumber \\
\sum_{l l'} \sqrt{(2l+1)(2l'+1)}
 \langle (\eta , 0)l;(\eta ,0)l || (\lambda_1,\mu_1) 10\rangle \\
 \langle (\eta' , 0)l';(\eta' ,0)l' || (\lambda_2,\mu_2) 10\rangle
\langle (\lambda_1,\mu_1) 10; (\lambda_2,\mu_2) 10 || (\lambda_0,\mu_0)
10 \rangle
\}.\end{eqnarray}

\newpage

{\large\bf Table Captions}\\

\bigskip\noindent
Table 1: Irreps and $C_2$ values for protons and neutrons in $^{159}$Tb.
Only the seven irreps with largest $C_2$ values are listed .

\bigskip\noindent
Table 2: The 15 pseudo SU(3) irreps used in the description of $^{159}$Tb
bands.

\bigskip\noindent
Table 3: Experimental B(E2) transition strengths for $^{159}$Tb.

\bigskip\noindent
Table 4: Explicit form of the irreps referred to as the components of the
wavefunctions in Figure 6.

\newpage


\begin{table*}[h]
\begin{center}
\begin{tabular}{cc|cc}
$(\lambda_\pi, \mu_\pi )$ & $C_2$ & $(\lambda_\nu, \mu_\nu )$ &
$C_2$ \\ \hline
(10,4)& 198 & (18,4) & 478 \\
(7,7) & 189 & (20,0) & 460 \\
(11,2)& 186 & (16,5) & 424 \\
(2,11)& 186 & (17,3) & 409 \\
(8,5) & 168 & (18,1) & 400 \\
(5,8) & 168 & (13,8) & 400 \\
(9,3) & 153 & (14,6) & 376 \\
\label{Table1}
\end{tabular}
\centerline{Table 1}
\end{center}

\bigskip

\begin{center}
\begin{tabular}{cc|cccccc}
$(\lambda_\pi, \mu_\pi )$ &$(\lambda_\nu, \mu_\nu )$ &
\multicolumn{6}{c}{total $(\lambda, \mu )$ } \\ \hline
(10,4) &(18,4) & (28,8) &(29,6) &(30,4)& (31,2) & (32,0) & (26,9)\\
(11,2) &(18,4) & (29,6) & (30,4) & (31,2) \\
(10,4) &(20,0) & (30,4) \\ 
(11,2) &(20,0) & (31,2) \\
(7,7)  &(18,4) & (25,11) & (26,9) \\
(10,4) &(16,5) & (26,9) \\
(8,5)  &(18,4) & (26,9)
\label{table2}
\end{tabular}
\centerline{Table 2}\end{center}
\end{table*}

\begin{table*}[h]
\begin{center}
\begin{tabular}{cc}
\multicolumn{2}{c}{Experimental B(E2) for $^{159}$Tb} \\
$J_\#^+ \rightarrow {(J-2)}_\#^+$ & Exp $[e^2 b^2\times 10^{-2}]$ \\ \hline
${7/2}^+_1 \rightarrow {3/2}^+_1$  & $73.7 \pm 10.2$ \\
${9/2}^+_1 \rightarrow {5/2}^+_1$  & $111.5 \pm 6.1$ \\
${11/2}^+_1 \rightarrow {7/2}^+_1$ & $147.9 \pm 3.6$ \\
${13/2}^+_1 \rightarrow {9/2}^+_1$ & $166.3 \pm 4.6$ \\
${15/2}^+_1 \rightarrow {11/2}^+_1$& $158.1 \pm 10.7$ \\
${17/2}^+_1 \rightarrow {13/2}^+_1$& $154.5 \pm 10.2$ \\
                                               &                  \\
${1/2}^+_1 \rightarrow {5/2}^+_1$  & $2.56$         \\
                                               &                  \\ \hline
$J_\#^+ \rightarrow {(J-1)}_\#^+$ & Exp $[e^2 b^2\times 10^{-2}]$ \\ \hline
${5/2}^+_1 \rightarrow {3/2}^+_1$   & $186.7 \pm 9.7$  \\
${7/2}^+_1 \rightarrow {5/2}^+_1$   & $117.7 \pm 25.6$ \\
${9/2}^+_1 \rightarrow {7/2}^+_1$   & $59.9  \pm 6.6$  \\
${11/2}^+_1 \rightarrow {9/2}^+_1$  & $57.3  \pm 5.6$  \\
${13/2}^+_1 \rightarrow {11/2}^+_1$ & $32.2  \pm 4.1$  \\
${15/2}^+_1 \rightarrow {13/2}^+_1$ & $38.4  \pm 6.6$  \\
						&		   \\
\end{tabular}
\centerline{Table 3}
\label{fig:table3}
\end{center}
\end{table*}

\begin{table*}[h]
\begin{center}
\begin{tabular}{c|c|cc}
\multicolumn{4}{c}{labels and irreps for $^{159}$Tb} \\
$label$ & ${(\lambda,\mu)}_T S_T$ & $(\lambda_\pi,\mu_\pi) S_\pi$ &
$(\lambda_\nu,\mu_\nu) S_\nu$ \\ \hline
${\dagger}$       & $(30,4){\frac 1 2}$ & $(10,4){\frac 1 2}$ & $(20,0)0
$\\
${\diamond}$      & $(28,8){\frac 1 2}$ & $(10,4){\frac 1 2}$ & $(18,4)0
$\\
${\star}$         & $(30,4){\frac 1 2}$ & $(11,2){\frac 1 2}$ & $(18,4)0
$\\
${\triangleleft}$ & $(31,2){\frac 1 2}$ & $(11,2){\frac 1 2}$ & $(20,0)0
$\\
${\bullet}$       & $(31,2){\frac 1 2}$ & $(10,4){\frac 1 2}$ & $(18,4)0
$\\
${\ast}$          & $(29,6){\frac 1 2}$ & $(11,2){\frac 1 2}$ & $(18,4)0
$\\
${\circ}$         & $(29,6){\frac 1 2}$ & $(10,4){\frac 1 2}$ & $(18,4)0
$\\
${\times}$        & $(30,4){\frac 1 2}$ & $(10,4){\frac 1 2}$ & $(18,4)0
$\\
${\triangleright}$& $(31,2){\frac 1 2}$ & $(11,2){\frac 1 2}$ & $(18,4)0
$\\
\multicolumn{4}{c}{} \\
\end{tabular}
\centerline{Table 4}
\label{fig:table4}
\end{center}
\end{table*}

\clearpage\newpage

{\large\bf Figure Captions} \\

\bigskip\noindent
Fig. 1: Experimental (left hand side) and theoretical (right hand side)
energies
for the first four bands in $^{159}$Tb.

\bigskip\noindent
Fig. 2: Experimental (left hand side) and theoretical (right hand side)
energies
calculated {\em without} single-particle energies.

\bigskip\noindent
Fig. 3: Energy bands in $^{159}$Tb. The first and tenth columns depict the
experimental spectra. The second column shows the spectra obtained using
Hamiltonian (\ref{ham}). The next seven columns show the same spectra obtained
using Hamiltonian (\ref{ham}) but with one term neglected, which is identified
below each column: the third column does not include pairing, the fourth column
does not include proton pairing, the fifth column does not include neutron
pairing,
the sixth column does not include the $K^2_J$ term, the seventh column does not
include the $J^2$ term, the  eighth eighth column does not include the
asymmetry
term, and finally, the ninth column does not include any of the three rotor
terms. Insert (a) shows the
${3/2}_1$ band, insert (b) the ${5/2}_1$ band and insert (c) the
${1/2}_1$ band.

\bigskip\noindent
Fig. 4: $^{159}$Tb theoretical energies and  $B\{E2;J \rightarrow (J-1) \}$
transition strengths (in units $e^2 b^2 \times 10^{-2}$). The latter are
written
close to the arrows indicating each transition.

\bigskip\noindent
Fig. 5: $^{159}$Tb theoretical energies and   $B\{E2;J \rightarrow (J-2) \}$
transition strengths, with the same convention as for Fig. 4.

\bigskip\noindent
Fig. 6: SU(3) components of the calculated eigenstates in the first four
bands of $^{159}$Tb. The angular momentum is listed in the horizontal axis. 
The symbols refer to the first column in Table 4. Insert (a) shows  the 
${3/2}_1$ band, insert (b) the ${5/2}_1$ band, insert (c) the
${1/2}_1$ band and insert (d) the ${3/2}_2$ band.

\newpage

\begin{figure*}[h]
\epsfxsize=9.5in
\centerline{\epsfbox{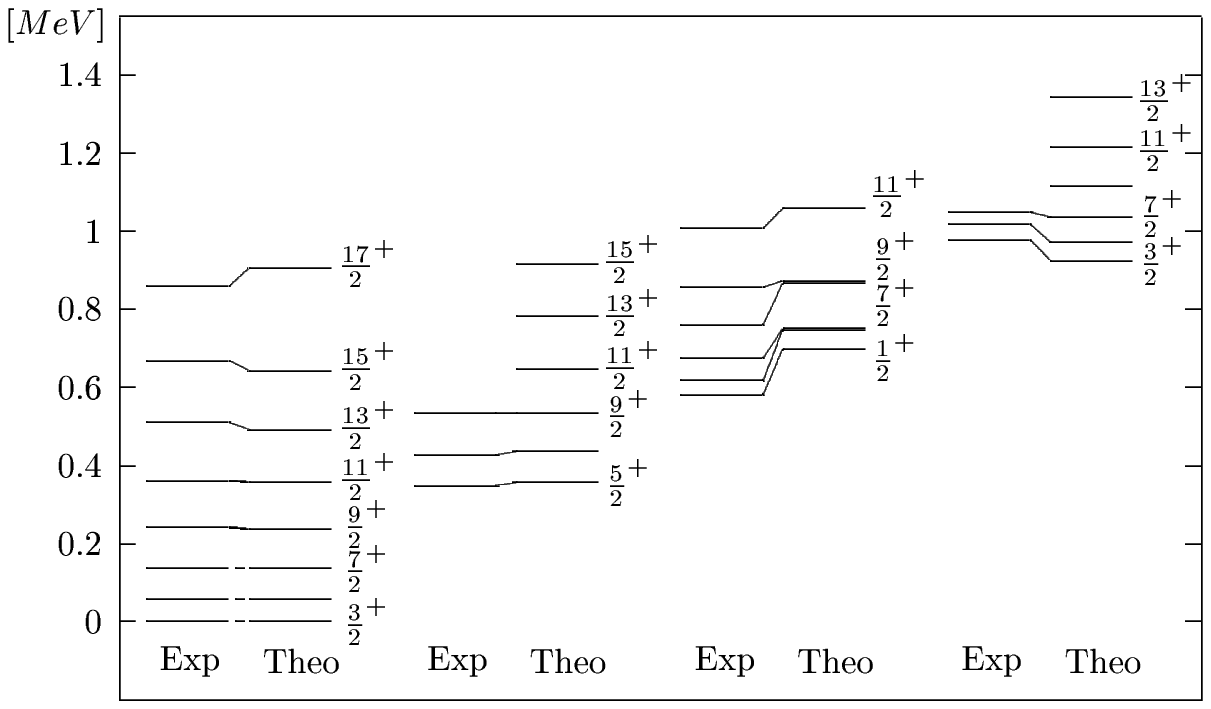}}
\vspace*{-17cm}
\centerline{Fig. 1}
\end{figure*}

\begin{figure*}[h]
\vspace*{-5cm}
\epsfxsize=9.5in
\centerline{\epsfbox{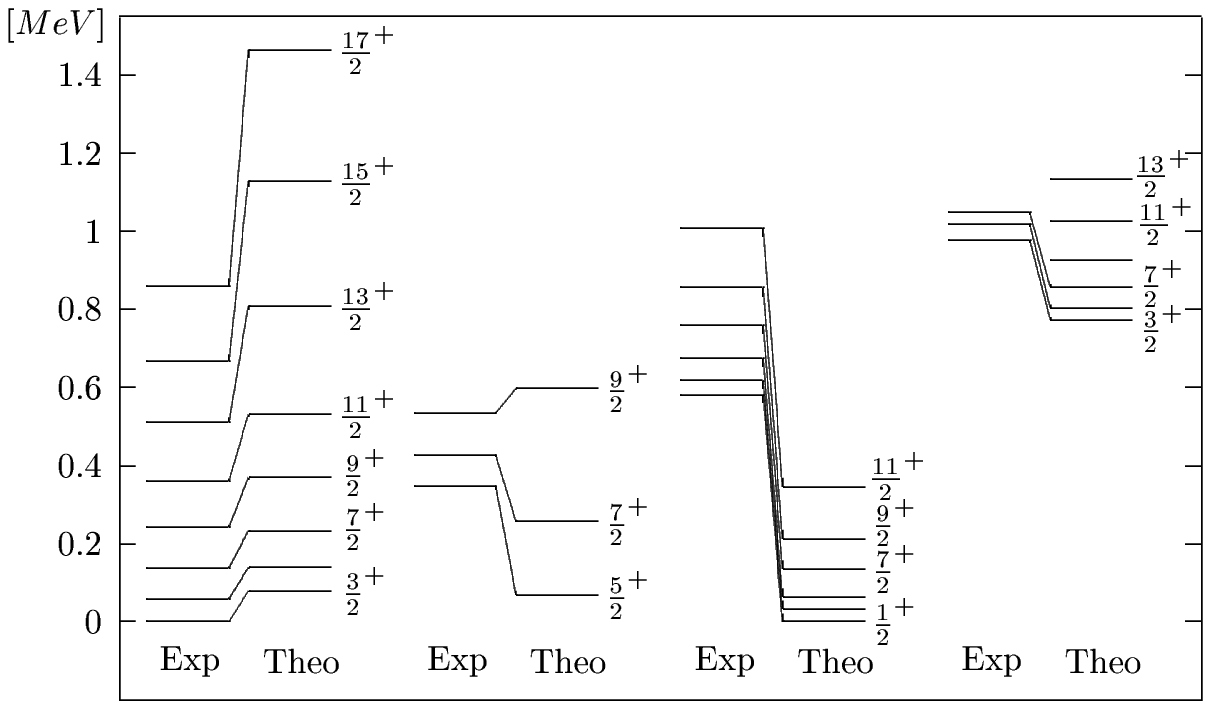}}
\vspace*{-17cm}
\centerline{Fig. 2}
\end{figure*}

\clearpage\newpage
\begin{figure*}[h]

\vspace*{-10cm}
\epsfxsize=9.5in
\centerline{\epsfbox{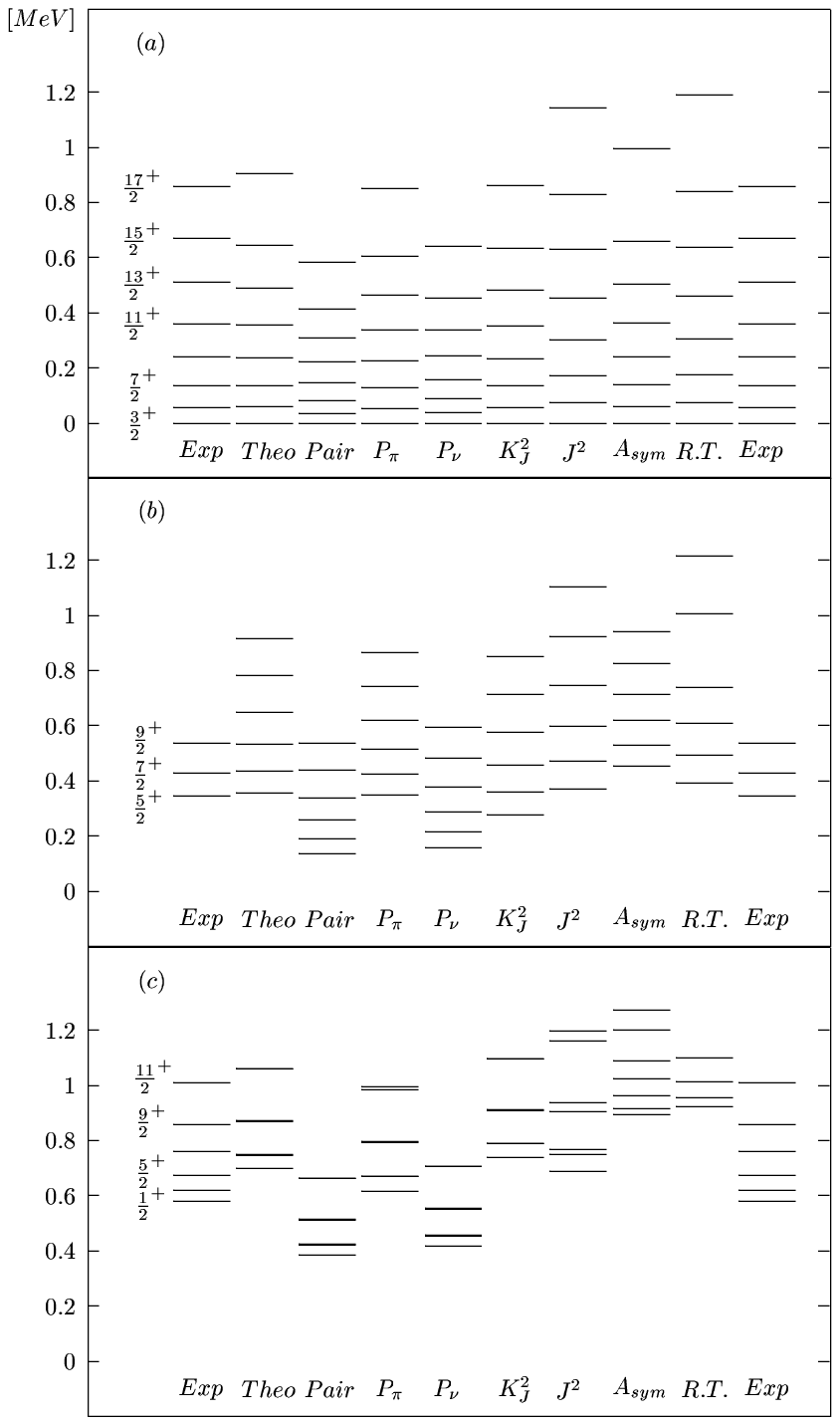}}
\vspace*{-7cm}
\centerline{ Fig. 3}
\end{figure*}

\begin{figure*}[h]
\vspace*{-7cm}
\epsfxsize=6.5in
\centerline{\epsfbox{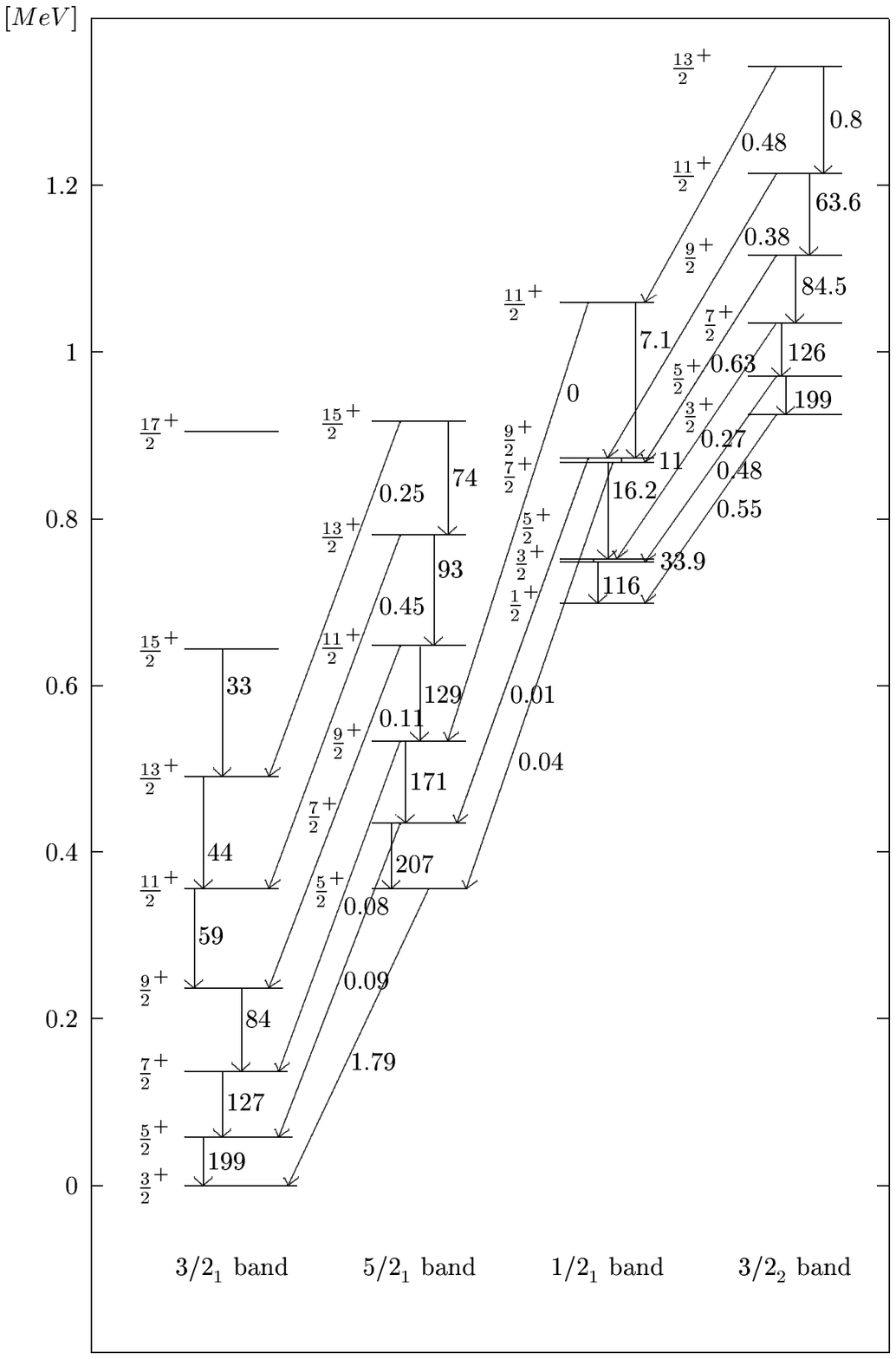}}
\vspace*{-3cm}
\centerline{Fig. 4}
\label{Figure4}
\end{figure*}

\clearpage\newpage
\begin{figure*}[h]
\vspace*{-3cm}
\epsfxsize=6.5in
\centerline{\epsfbox{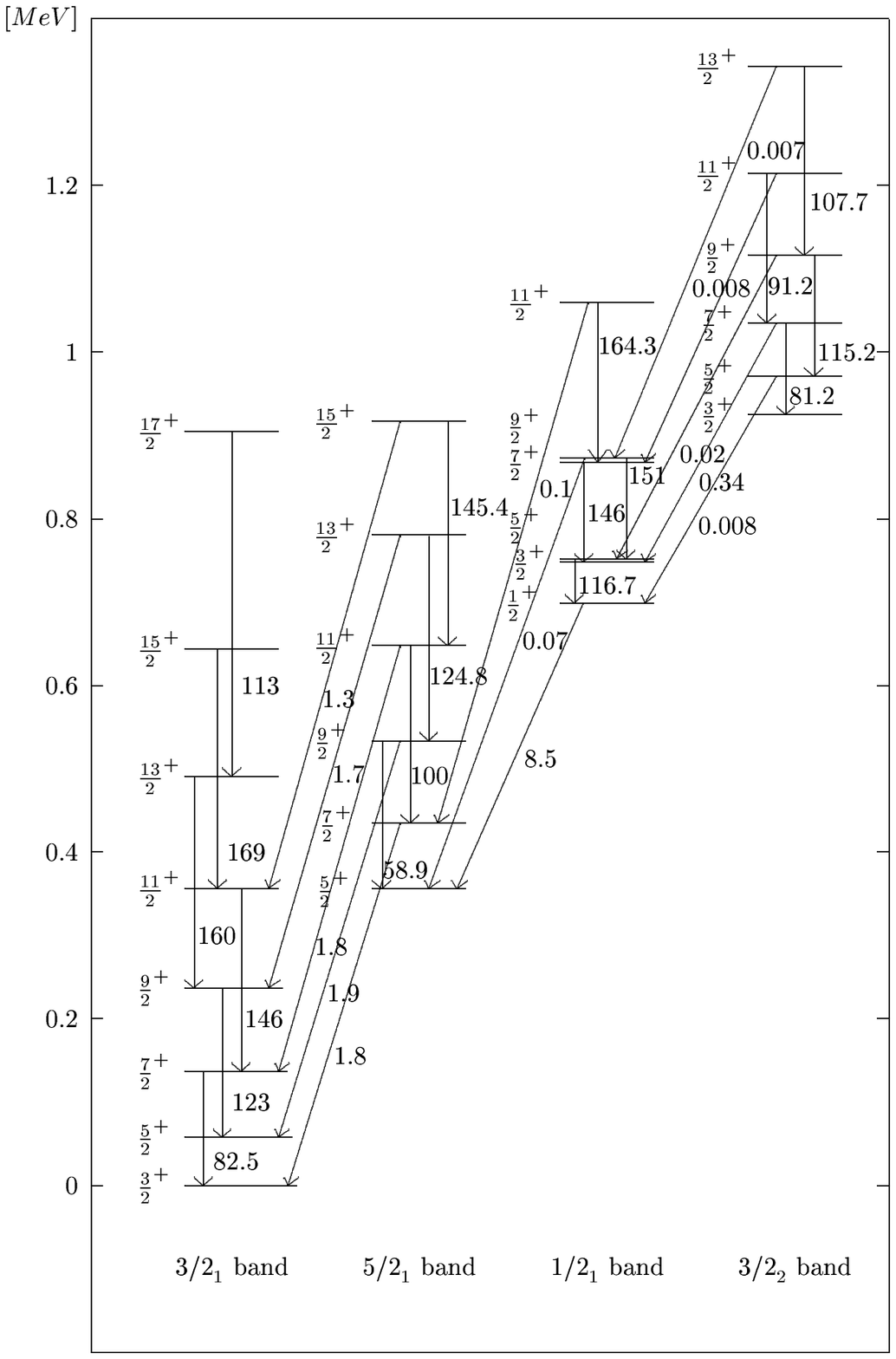}}
\vspace*{-3cm}
\centerline{Fig. 5}
\label{Figure5}
\end{figure*}

\clearpage\newpage
\begin{figure*}[h]
\vspace*{-5cm}
\epsfxsize=9.5in
\centerline{\epsfbox{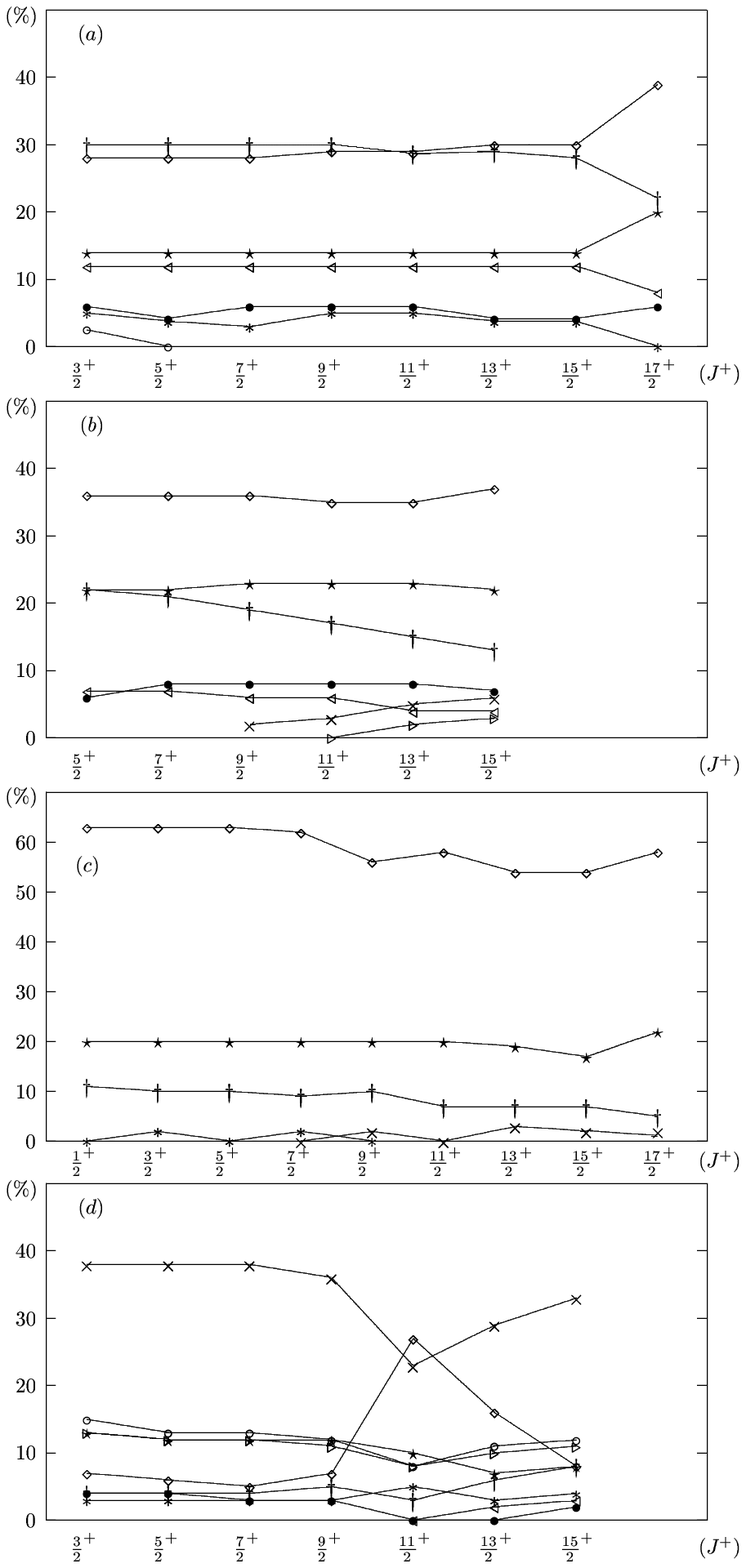}}
\vspace*{-7cm}
\centerline{Fig. 6}
\end{figure*}

\end{document}